\documentclass[12pt]{iopart}
\usepackage{iopams}
\usepackage{graphicx}
\usepackage{cite}
\usepackage{psfrag}
\usepackage{dcolumn}
\usepackage{latexsym}
\usepackage{amssymb,latexsym,mathrsfs,color}
\usepackage{amsfonts}

\newcommand{\beq}{\begin{equation}}
\newcommand{\eeq}{\end{equation}}
\newcommand{\beqn}{\begin{eqnarray}}
\newcommand{\eeqn}{\end{eqnarray}}
\newcommand{\bearr}{\begin{array}}
\newcommand{\enarr}{\end{array}}

\def\bea{\begin{eqnarray}}
\def\eea{\end{eqnarray}}
\def\ba{\begin{array}}
\def\ea{\end{array}}

\def \be{\begin{equation}}
\def \ee{\end{equation}}
\def \bc{\begin{center}}
\def \ec{\end{center}}
\def \aa {\alpha_{_A}}
\def \ab {\alpha_{_B}}
\def \ac {\alpha_{_C}}
\def \ba {\beta_{_A}}
\def \bb {\beta_{_B}}
\def \bc {\beta_{_C}}
\def \tra {\theta_{RA}}
\def \trb {\theta_{RB}}
\def \tr {\theta_R}
\def \rd{\textcolor{black}}

\begin{document}

\title{Asymmetric exclusion processes on a closed network with bottlenecks}
\author{Rakesh Chatterjee, Anjan Kumar Chandra and Abhik Basu }
\ead{rakesh.chatterjee@saha.ac.in, anjanphys@gmail.com, abhik.basu@saha.ac.in}
\address{Condensed Matter Physics Division, Saha Institute of
Nuclear Physics, Kolkata 700064, India}

\begin{abstract}
We  study the generic nonequilibrium steady states in asymmetric exclusion
processes on a closed network with bottlenecks. To this end we proposes and 
study closed simple networks with multiply-connected non-identical
junctions. Depending upon the parameters that define
the network junctions and the particle number density, the models display 
phase transitions with both static and moving density inhomogeneities. 
The currents in the models can be
tuned by the junction parameters. Our models highlight how extended
and point defects may affect the density profiles in a closed
directed network. Phenomenological implications of our results are discussed.
\end{abstract}


\maketitle

\section{Introduction}

Simplest physical modeling of \rd{directed or active} classical transports in one
dimensions (1D) {\rd{(e.g., along narrow channels)}} are often provided in terms of asymmetric simple exclusion processes. A well-known example of directed transport in
1D is the system of unidirectionally moving
vehicular traffic along roads~\cite{dcrev,helbing}. Narrow roads
with excluded volume interactions between vehicles having no
possibility of overtaking, are well-described by the totally
asymmetric simple exclusion process (TASEP)~\cite{tasep-rev}.
Important phenomenological questions about traffic networks include how defects (both
extended and point) along a road may control the density profiles
and currents in the steady states. We concern ourselves in exploring
these issues in terms of a simple model in this paper.

In order to focus on the essential physics of the system, we
consider a minimal closed directed network (hereafter Model I) consisting of just three
segments, with each executing TASEP, two of them (marked $T_A$ and
$T_B$ in Fig.~\ref{fig:model1}) being parallelly and the third one
($T_C$) anti-parallelly attached to the multiply connected left
($LJ$) and right ($RJ$) junctions. Junctions $LJ$ and $RJ$  are {\em
non-identical}, since the system is {\em directed}. Notice that
$T_C$, the antiparallel TASEP channel may be viewed as an extended
defect or \rd{an extended bottleneck} in the system, since its maximum current carrying capacity is {\em less} than the total maximum current carrying capacities of the
two parallel channels. Thus, $T_C$ {\em restricts} the maximum
permissible system current. Therefore, $T_C$ acts like an (extended)
bottleneck in the system. While in general the length of an extended
defect is not necessarily same as the lengths of either of the
parallel channels, but for simplicity we take all of them to have
the same length with equal number of lattice sites. {Although the
branching of particles at $LJ$ is controlled by a parameter
$\theta_L$, at $RJ$ there is no analogous control parameter.}

  There have been a number of applications of
TASEP and TASEP-like models in studies on directed vehicular traffic
along roads, studying various physical aspects of  traffic jam described as TASEP or TASEP-like systems. For instance, the Nagel-Schreckenberg model, closely related to
TASEP, has been proposed as a theoretical description for freeway
traffic. The model displays traffic jam when the vehicle density is
high~\cite{nagel}. Shock propagation in traffic systems has also
been studied using TASEP \cite{Domany,Schutz}. In a recent work,
Ref.~\cite{Ito} has studied interactions between vehicles and
pedestrians in terms of a TASEP with a bottleneck at a boundary
caused by interactions. Furthermore, Ref.~\cite{zhang} has
considered an ``optimal velocity model" with two kinds of vehicles
(fast and slow). With specific lane change rules and under periodic
boundary conditions, the traffic states change with increasing
densities. Further, Ref.~\cite{zhang} finds a new phenomenon of
ratio inversion. In a 1D traffic model, Ref.~\cite{zhang2} studies
the impact of disruptions on road networks, and the recovery process
after the disruption is removed from the system. Extensive reviews
on applications of TASEP-like models in the studies of vehicular
traffic and related areas are available in
Refs.~\cite{dcrev,helbing}. Our studies on Model I are complementary to these
model studies. We have strict particle number conservation and an
extended defect. Thus Model I should be useful for studying
traffic problems where  the vehicle current is controlled by an
extended defect for a fixed number of vehicles. In the
\ref{model2} of this article we discuss an extension of
Model I, by including additional junction parameters $\tra$ and
$\trb$ that control the relative flow of particles from $T_A$ and
$T_B$ to $T_C$ at the junction $RJ$, which appear as point defects
in the model, \rd{can be considered as traffic signals at that junction}. 
We call this Model II; see Fig.~\ref{fig:model2}. This further
allows us to find how the steady state densities are
affected by the point defects at $RJ$. Our results reveal
interesting interplay between the extended and point defects, which
ultimately controls the steady state densities.

We now compare Model I with a recently proposed model by
us~\cite{rc_akc_ab}, where two asymmetric exclusion processes in the
form of two TASEPs are coupled with $1D$ diffusive motion or SEP.
The principal
physical difference between  them is that, the diffusive channel
in Ref.~\cite{rc_akc_ab} does {\em not} control the maximum current
in the model, allowing each of the TASEP channels there to reach
their individual maximum currents. In contrast, the maximum current
in the present models is {\em limited} by $T_C,$ the
extended bottleneck, which is absent in Ref.~\cite{rc_akc_ab}. \rd{As we shall see below, the extended bottleneck has major consequences on the phases and density profiles in Model I; some phases individually accessible to an open TASEP or accessed by the model of Ref.~\cite{rc_akc_ab} are ruled out here.}

\rd{Here is a summary of our specific results. Both Model I and Model II display generic 
nonequilibrium phase transitions associated with
a variety of density profiles ranging from uniform (flat) profiles
to localised (LDW) or static and delocalised (DDW) or moving domain walls, controlled by
the particle number and the junction parameters. Nonetheless, there are 
significant differences between the density profiles of Model I and Model II. 
Due to the absence of any point defect at $RJ$, the DDWs in $T_A$ and $T_B$ in 
Model I are always {\em overlapping}, a consequence of a special symmetry at the 
delocalisation point; in contrast, the DDWs {\em do not overlap} in Model II. 
Furthermore, the presence of the point defect at $RJ$ in Model II allows for a 
class of phases, which are not permitted in Model I. For example, in Model II, 
the extended bottleneck $T_C$ can be in its LD-HD coexistence or LDW phase, 
whereas such a possibility is ruled out in Model I on the ground of current 
conservation and symmetry. On the whole, our models here serve as good 
candidates to study the interplay between extended and point defects, and number 
conservation in closed, simple, directed networks. Our calculational 
framework may be systematically extended to a network with a larger number of 
segments/joints. Despite the simplicity and the minimalist nature of our 
models, the above results should be potentially relevant in the 
contexts of
defect-controlled vehicular traffic in closed network of roads. The rest of the paper is organised as follows. In Sec.~\ref{model1} we construct Model I. In Sec.~\ref{profile} we obtain the steady state density profiles and the phase diagram of Model I. In Sec.~\ref{conclu} we conclude.
In \ref{model2} we
introduce Model II. The corresponding phase diagram and the density profiles are 
analysed, respectively, in
\ref{model2phase} and
\ref{model2density}.}

\section{Construction of Model I}\label{model1}
\begin{figure}[t]
\centering
\includegraphics[width=7.5cm]{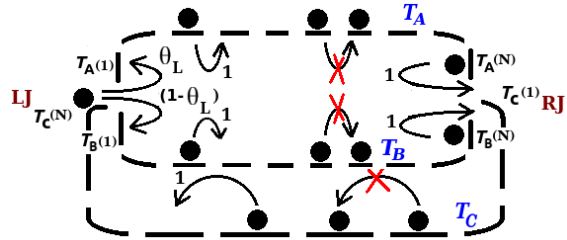}
\caption{(Color online) Schematic diagram of the Model I; $LJ$/$RJ$
refer to the left/right junctions. Site labels run from $i=1$ to $N$
from $LJ$ to $RJ$ for $T_{A,B}$ and for $T_C$ from $RJ$ to $LJ$.}
\label{fig:model1}
\end{figure}

In Model I, each channel consists of equal number of sites
designated by $i=1,2,...N.$ As shown in Fig.~\ref{fig:model1}, the
left (right) most site of $T_A$ is labeled as $T_A(1)$ ($T_A(N)$).
Similar labeling follow for $T_B$ and $T_C$. Particles from $T_C$
enter at the left end of $T_A$ ($T_B$) if empty with probability
$\theta_L$ ($1-\theta_L$) and exit to the right end of $T_C$ if
vacant from the right end of $T_A$ and $T_B$ with rate unity.
In each of $T_A$, $T_B$ ($T_C$) particles can only hop with rate
unity to the right (left) neighbor if it is empty. The global
particle density is $n_p = N_p/3N$, with $N_p$ being the total
particle number in the system. In this model, the phases are
parametrized and tunable by $n_p$ and $\theta_L.$ The presence of
$LJ$ and $RJ$ breaks the translational invariance, and hence,
non-trivial steady states are expected~\cite{hinsch}.

\section{Steady state density profiles}
\label{profile}
We use mean-field theory (MFT) together with extensive Monte-Carlo
simulation (MCS) by using random sequential updates to
obtain the steady state density profiles in our model. In the MFT,
the system is considered as a collection of three separate TASEP
channels with effective entry and exit rates~\cite{hinsch,effective}
($\alpha_m$ and $\beta_m$, respectively for channel
$T_m,\,m=A,B,C$), to be determined by applying the condition of
constancy of particle currents at the junctions $LJ$ and $RJ$ and in
the bulk. These immediately allow us to obtain the phases of
the individual channels and hence of the whole model in terms of
the known results for TASEP with open boundaries.
In the discrete lattice description of
our model, we denote the density at a particular site $i$ in channel
$T_m$ by $\rho_m^i=\langle n_m^i \rangle$, $m=A,B,C$, {$n_m^i = 0$
or $1$ and $\langle...\rangle$ denotes time and configuration
averages}, whereas in MFT considering continuum limit the density is
defined as $\rho_m(x)$, where $x=i/N,$ and in the thermodynamic
limit (TL) $N\gg 1,$ $x$ in the range $0\leq x\leq1.$ Given the
symmetry between $T_A$ and $T_B$, $\rho_A(x)$ and $\rho_B(x)$
interchange when $\theta_L$ interchanges with ($1-\theta_L$).
This symmetry is identical to the one in the model of
Ref.~\cite{rc_akc_ab}.

Recall that an isolated TASEP can be in four different phases in its
steady states: Low density (LD), high density (HD), maximal current
(MC) and coexistence or domain wall (DW) phases~\cite{tasep,krug};
clearly there are almost 64 possibilities for the overall
density profiles of our model.  For the
ease of notation and compactness, we denote a phase by (X-Y-Z)
where X/Y/Z refers to the phase of $T_A/T_B/T_C$ for a given
choice of $n_p$ and $\theta_L.$ Obviously, when $n_p$ is very low,
all of $T_A,T_B,T_C$ will be in their LD phases for any values of
$\theta_L$ and thus the system adopts the phase LD-LD-LD. Similarly,
for a very high $n_p$ all the channels will be in their HD phases
and the system is in its HD-HD-HD phase. We now discuss the
admissibility of the intermediate phases as $n_p$ varies.
Bulk current conservation in the steady states,
for any value of $\theta_L$, yields (in the MFT),
\begin{equation}
\rho_A(x)[1-\rho_A(x)] + \rho_B(x) [1-\rho_B(x)] = \rho_C(x)
[1-\rho_C(x)]. \label{currbulkcon}
\end{equation}
The maximum of the right hand side of Eq.~(\ref{currbulkcon}) is $1/4,$ corresponding
to the MC phase in $T_C.$ This immediately rules out MC phases in
$T_A$ or $T_B$ for $\theta_L\neq 0,1$. Thus, phases ($X,MC,Z$) or ($MC,Y,Z$) are not
allowed. In addition, phases LD-LD-HD and HD-HD-LD are prohibited
for any $\theta_L \neq 0$ or $1.$ If $T_C$ is in its MC phase for
a given $(n_p,\theta_L)$ value, increasing $n_p$ will
lead to addition of more particles with no change in
$\rho_C(x)=1/2$ in the bulk. Hence, the extra particles should accumulate
in either $T_A$ or $T_B$ or both, without any change in the total current given by
Eq.~(\ref{currbulkcon}). \rd{Hence, any addition of particles is then expected to manifest in the form of LDWs in $T_A$ or $T_B$ or both, which leave the currents unchanged.} Our detailed steady state density profiles confirm this physically intuitive expectation.

\begin{figure}[h]
\centering
\includegraphics[width=9.5cm]{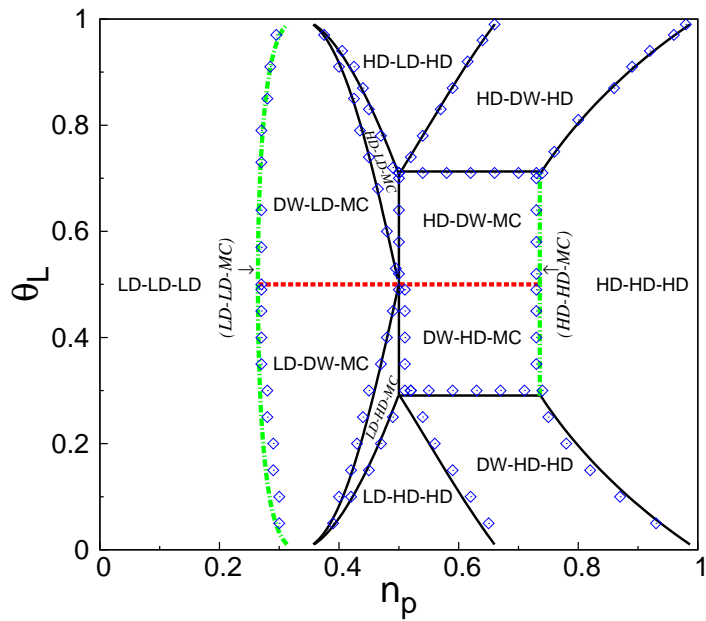}
\caption{(Color online) Phase diagram in $(n_p,\theta_L)$ -plane for
Model I, phase boundaries are represented by black continuous lines,
the line-phases are denoted by green dashed-dotted lines and DDW
appears on red dotted line as obtained from MFT, whereas
corresponding data points are from MCS with $N = 500.$}
\label{phase_mft}
\end{figure}

The results from Model I are summarized in the phase diagram
Fig.~\ref{phase_mft} where different phases are marked in the
$(n_p,\theta_L)$ -plane.
The phase diagram is quite complex in having a large number of
phases, as expected. Notice that it is symmetric
about the line $\theta_L=1/2$, a consequence of the interchange
between $T_A$ and $T_B$ with an interchange between $\theta_L$ and
$(1-\theta_L)$. Notice that in (\ref{phase_mft}) some of the phases are represented by finite areas, whereas others appear just as lines. We discuss the physical principles for
the calculation of the density profiles below with some illustrative
examples. \rd{Full calculations of the steady state densities $\rho_A,\,\rho_B,\,\rho_C$ for Model II are given in Appendix as reference.}

\begin{figure}[h]
\centering
\includegraphics[width=7.5cm]{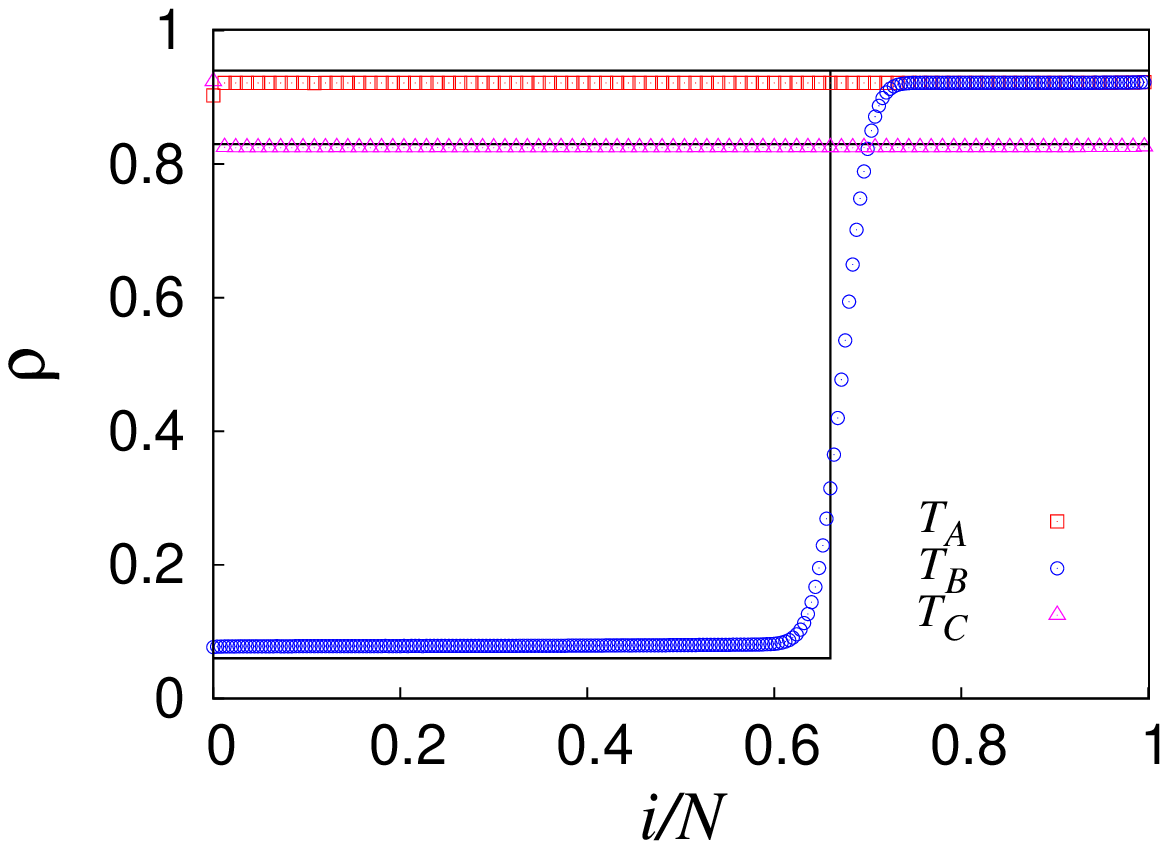}\includegraphics[width=7.5cm]{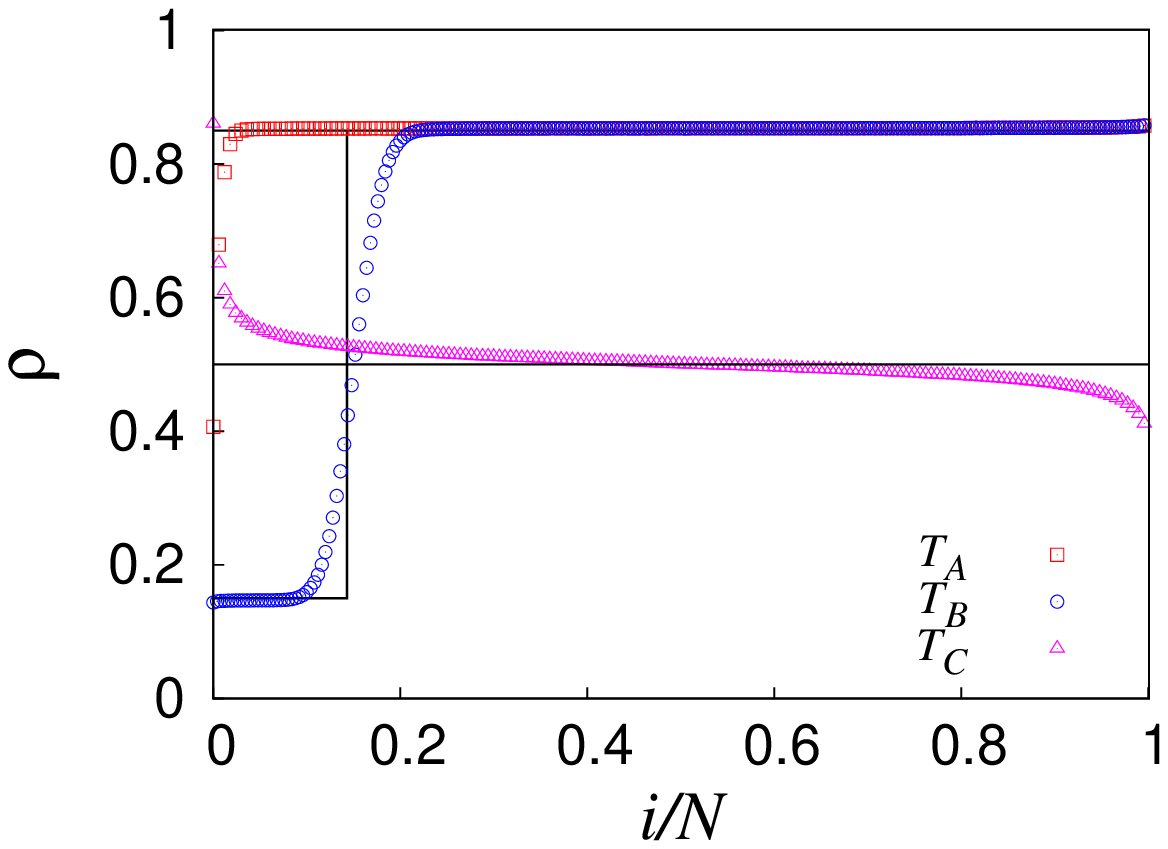}
\caption{(Color online) (left) Density profiles for $T_A,T_B,T_C$
obtained with $n_p = 0.70$, $\theta_L = 0.90$ and $N = 500,$
displaying (HD-DW-HD) phase. LDW is found at $x_B^w=0.661$ from MFT which matches with the numerical value $0.672.$ 
(right) Density profiles for (HD-DW-MC) phase
obtained with $n_p = 0.70$, $\theta_L = 0.55$ and $N = 500.$
LDW is found at $x_B^w=0.143$ from MFT which matches with the numerical value
$0.149.$ Numerical data displayed by points whereas solid lines denote MFT results.}
\label{fig:hddwhd-mc}
\end{figure}

Consider the phase HD-DW-HD, with an LDW in $T_B$ at $x_B^w$ as shown in Fig.~\ref{fig:hddwhd-mc}.
Since $T_A$ and $T_C$ are in their HD phases, their densities are
$\rho_A=1-\ba$ and $\rho_C=1-\bc$, neglecting the boundary layers
(BL). At $RJ$, both $T_A$ and $T_B$ have no BL; current
conservation at $RJ$ yields
$\ba = \bb.$ Noting that there is an LDW in $T_B$,  current
conservation at $LJ$ gives, 
\be 
\bb(1-\bb) = (1-\bc)(1-\theta_L)\bb. 
\label{eq:hddwhd-1} 
\ee
Overall current conservation in the bulk gives,
\begin{equation}
\ba(1-\ba) + \bb(1-\bb) = \bc(1-\bc).
\label{eq:hddwhd-2}
\end{equation}
Using particle number conservation and neglecting BLs in TL we have,
\begin{equation}
3n_p = 1-\ba + \bb + (1-x_B^w)(1-2\bb) + 1-\bc .
\label{eq:hddwhd-4}
\end{equation}
The boundary lines between
HD-LD-HD and HD-DW-HD phases and HD-DW-HD and HD-HD-HD phases may be obtained
respectively by setting $x_B^w=1$ and $x_B^w=0;$
\begin{eqnarray}
3n_p &=& 3-\ba - \bb - \bc, \nonumber \\
3n_p &=& 2 - \ba + \bb - \bc.
\label{eq:hddwhd-5}
\end{eqnarray}
However, these boundary lines do not span over the entire range 
$0\leq\theta_L\leq 1$, but get cut off at
 $\theta_L<1$ at which another phase HD-DW-MC appears; see Fig.~\ref{fig:hddwhd-mc} for the corresponding density profiles.
The phase boundary  between HD-DW-HD and HD-DW-MC is obtained by setting $\bc=1/2.$
Now putting that in Eqs.~(\ref{eq:hddwhd-1}) and (\ref{eq:hddwhd-2}) we have the equation of the horizontal boundary line as $\theta_L=1/\sqrt{2}.$

The other boundaries of  HD-DW-MC phase may be obtained as follows. Current conservation at the $RJ$ gives $\ba=\bb.$ Again from overall current in this phase yields
\be
\ba(1-\ba) + \bb(1-\bb) = 1/4.
\label{eq:hddwmc-1}
\ee
Considering $T_A$ in its HD phase, we get $\ba=0.146=\bb$ from
Eq.~(\ref{eq:hddwmc-1}).
%
Following the calculation logic outside above, the boundary lines between
HD-LD-MC and HD-DW-MC phases and HD-DW-MC and HD-HD-MC phases obtain respectively as
\begin{eqnarray}
3n_p &=& 3/2,  \nonumber \\
3n_p &=& 5/2-2\ba. \label{eq:hddwmc-3}
\end{eqnarray}
The phase HD-DW-MC 
Again,  HD-DW-MC does
not span over the whole range of $\theta_L$ but gets cut by another
phase DW-HD-MC as shown in the phase diagram.

In the special case with $\theta_L=1/2$, 
$\rho_A$ and $\rho_B$ should be statistically symmetric. Therefore, if one
of them satisfies the condition for an LDW, the other also must
satisfy the same. Thus, we should find a pair of domain walls one
each in $T_A$ and $T_B$, at locations $x_w^A$ and $x_w^B$,
respectively.
If there are indeed two LDWs in $T_A$ and $T_B$, then
$\aa=\ba=\ab=\bb.$ Furthermore, in this case $T_C$
must be in its MC phase, since once the LDWs are formed in
$T_A/T_B$, any addition of particles in the system will lead to
shifting of the LDW positions keeping the currents same.  Using
current conservation in the bulk,  we have,
\be
\aa(1-\aa) + \ab(1-\ab) = 1/4,
\label{eq:ddw-1}
\ee
solving which we have $\aa=0.146=\ab.$

\begin{figure}[h]
\centering
\includegraphics[width=7.5cm]{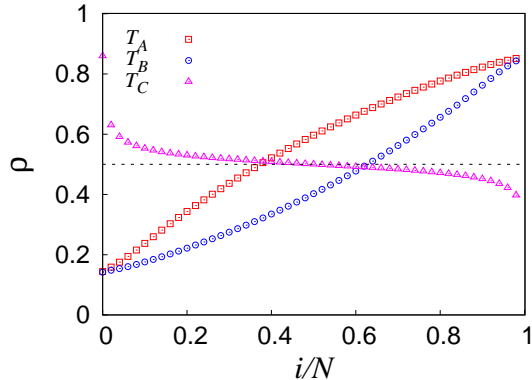}
\caption{(Color online) Density profiles for $T_A,T_B,T_C$
obtained numerically with $\theta_L = 0.50,$ $n_p=0.40$ and $N = 500,$
displaying DDW's in $T_A$ and $T_B$ with MC phase in $T_C.$}
\label{fig:ddw}
\end{figure}

Now from particle number conservation we get,
\be
3n_p=\aa + (1-x_A^w)(1-2\aa) + \ab + (1-x_B^w)(1-2\ab) + 1/2.
\label{eq:ddw-2}
\ee
Thus, the LDW positions $x_A^w$ and $x_B^w$ can no longer be determined uniquely. Since,
all (pairwise)
values of $x_A^w$ and $x_B^w$ that satisfy Eq.~(\ref{eq:ddw-2}), LDWs at each of such pairs of solutions
for $x_A^w$ and $x_B^w$ are physically valid solutions for the density profiles in $T_A$ and $T_B$.
In course of time the system should display all these
solutions, over long time averages $\rho_A$ and $\rho_B$ will
essentially appear as two inclined lines, which are the envelopes of
the allowed LDW solutions. In other words, we will observe two
DDWs. Our MFT analysis and physical arguments for DDWs are verified
by our MCS studies. Figure~\ref{phase_mft} shows the DDW-DDW-MC
line (red dotted line) as a borderline between the
DW-LD-MC and LD-DW-MC phases. In addition, representative DDW profiles for
$\rho_A$ and $\rho_B$ are shown in Fig.~\ref{fig:ddw}.

Two DDWs instead
of LDWs for $\theta_L=1/2$, can be argued from particle number conservation. Notice that for an LDW, its position is uniquely determined by
the particle number conservation; see, e.g., Eq.~(\ref{eq:hddwhd-4})
that gives $x_B^w$, the position of the LDW in $T_B$. However, if
there are two LDWs in the system, it is clear that an arbitrary
shift in the position of one of the LDWs, together with a
compensating reverse shift of the position of the second LDW keeps
the total particle number conserved. Thus, the LDW positions are
{\em not} uniquely determined. Hence the model displays DDWs. This
also explains why for a single LDW, its position in the steady state
is a quantity that remains stable by the dynamics of the system,
where as for DDWs, it is the {\em sum} of their positions that
remain stable.

Understandably, for a very low (high)
$\theta_L$, $T_A$ ($T_B$) is in its LD phase and $\rho_B >(<)\rho_A$
in general. For a fixed $\theta_L.$ when  $n_p$ is very
low, all of $T_A,T_B,T_C$ are in the LD phases, regardless of the
value of $\theta_L$. As $n_p$ increases, $T_B$ and $T_C$ move to
their DW/HD phase and $T_C$ to MC phases, while $T_A$ remains in its
LD phase for small $\theta_L$. This is due to the fact that for a
small $\theta_L,$ very few particles enter $T_A$ in comparison with
$T_B,$ regardless of $n_p.$ Eventually, as $n_p$ approaches unity
(the system is nearly filled), all of $T_A,T_B,T_C$ should be in
their HD phases, for any $\theta_L$. In general, the densities for
$T_A,T_B,T_C$ always change continuously across the phase
boundaries. Hence, with channel densities as the order parameter,
the transitions are second order in nature.

The phases LD-LD-MC and HD-HD-MC are just lines as represented
by green dashed-dotted lines in Fig.~\ref{phase_mft}. The fact that
they do not cover any area in the $(n_p,\theta_L)$ -plane can be
understood in simple physical terms. Since for these phases, $T_C$
is in MC phase, $\rho_C=1/2$ and the current through it is $1/4.$
Thus any putative change in $n_p$ is to be reflected by appropriate
changes in $\rho_A$ and $\rho_B$. Since $T_A$ and $T_B$ are assumed
to have uniform densities (LD or HD phases), any change in $n_p$
automatically leads to changes in $\rho_A$ and $\rho_B$ with
associated changes in their currents as well. This in turn spoils
the bulk current conservation, as the sum of their currents must be
$1/4$ (= current in $T_C$). Thus, these particular phases can be
realized only for one value of $n_p$ for a given
$\theta_L$, which explains why they appear as lines.
For instance, when both $T_A$ and $T_B$ are in their LD phases with densities
$\aa$ and $\ab$, respectively, with
$\aa/(1-\aa)=\theta_L/(1-\theta_L)$, the bulk currents here
satisfy the same Eq.~\ref{eq:ddw-1}.
Now, if $n_p$ is changed, say increased, then both $\aa$ and
$\ab$ rise keeping their ratio unchanged. This,
however, will spoil Eq.~\ref{eq:ddw-1}. Hence, these phases can survive
only for one particular value of $n_p$ for a given $\theta_L$, which
is consistent with their appearance as a line in phase diagram
Fig.~\ref{phase_mft}.

\section{Summary and outlook}\label{conclu}

To summarise, Model I reveals interesting interplay between
multiple links connecting non-identical junctions and the junctions
themselves, that determines the resulting macroscopic steady states
density profiles of the overall closed system. Since  $T_C$ is effectively an extended bottleneck in the system, our results in fact show the role of extended bottleneck in controlling the phases. Our scheme of MFT for obtaining
the steady state density profiles by using current and total
particle number conservations are generic and may be extended in
straightforward ways to more complex closed systems having larger
number of junctions and branches. As discussed in Sec.~\ref{model2} 
below, an additional point defect in Model II brings in new macroscopic  
steady state behaviour, including the possibility of an LDW in $T_C$ and 
non-overlapping DDWs in $T_A$ and $T_B$. The steady state bulk current in 
each of the segments may be easily determined from the knowledge of the 
corresponding densities; thus we show 
how the steady state currents may be controlled by the extended and point 
bottlenecks. Useful modifications in the
context of vehicular transport includes unequal lengths of the channels, allowing unequal channel lengths and different
velocities for different particles and/or along different channels,
impurities on the tracks and possibilities of change in speeds
(acceleration and braking)~\cite{trafficnew}.  Other possible relevant
extensions of theoretical interests include (i) one or some of the
branches allow bidirectional motion~\cite{rc_akc_ab,muhuri},(ii)
when there are local particle non conserving processes, e.g., random
attachment and detachment of particles~\cite{lktasep}, and (iii) the
presence of active and inactive agents (particles)~\cite{gov}. We
hope our work will motivate further works along these lines.

\vspace{1cm}

\appendix
\renewcommand{\theequation}{A-\arabic{equation}}
\setcounter{equation}{0}

\section{Model II}
\label{model2}
\begin{figure}[h]
\centering
\includegraphics[width=7.5cm]{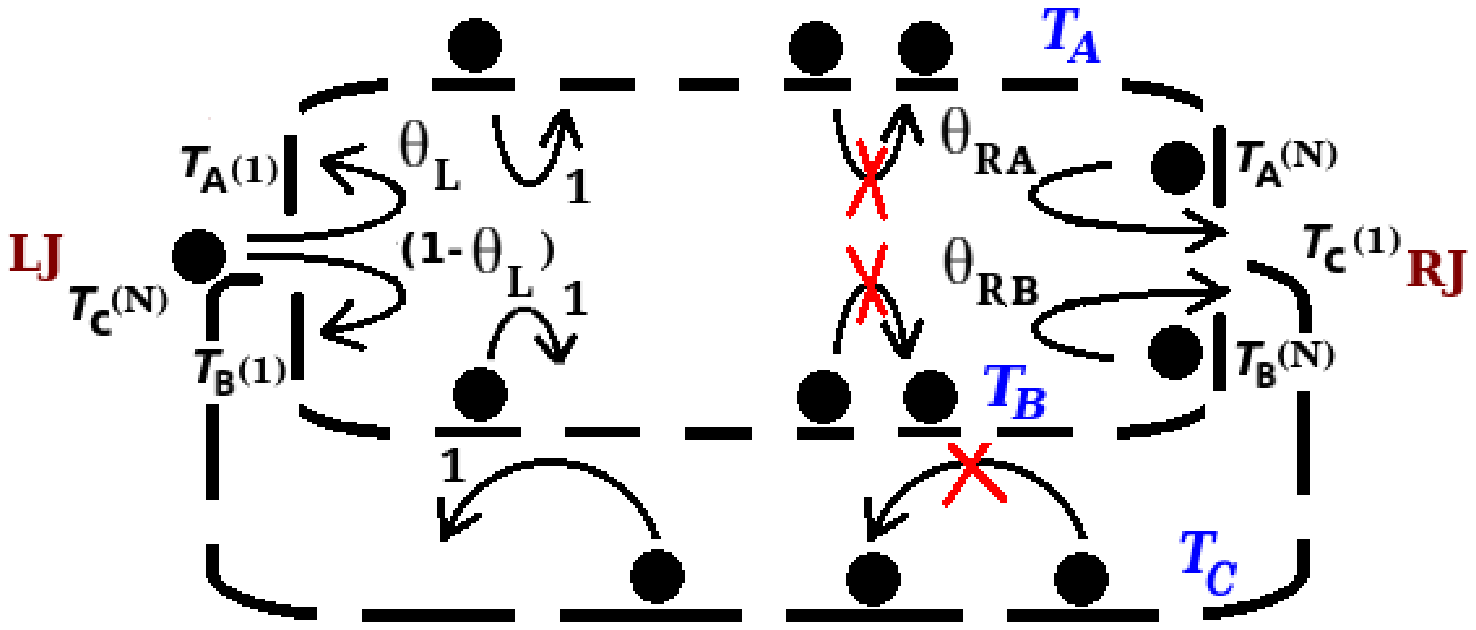}
\caption{(Color online) Schematic diagram of Model II with two
additional parameters $\tra$ and $\trb$ at the $RJ.$}
\label{fig:model2}
\end{figure}

We now introduce Model II, a generalization of Model I, that now includes a point
defect at the $RJ$, defined by  two new parameters $\tra$ and $\trb$; see Fig.~\ref{fig:model2}.
Here, $\theta_{RA}$ and $\theta_{RB}$  control the effective
particle flow into the the two channels $T_A$ and $T_B$ as point
defects at the junction position of regular TASEP channels.
Obviously, if we set $\tra = 1 = \trb,$ Model II should reduce
structurally to Model I. Analysis of the phases in Model II and
their differences with their counterparts in Model I allows us to
draw conclusions about the effects of the point defects at $RJ.$
 As a model for vehicular traffic
in a network of roads, these may potentially model dynamic obstacles (\rd{e.g., a traffic signal or sudden pedestrian crossings})
at $RJ$ in a simple way. We make a detailed comparison between Model
I and Model II in Appendix~\ref{apn:compare}.  To reduce the number of tunable
parameters, we set $\tra=1-\trb = \tr$ (say), and thereafter all results
 for Model II are obtained keeping this relation. 

\section{Phase diagram for Model II}\label{model2phase}
\begin{figure}[h]
\centering
\includegraphics[width=9.5cm]{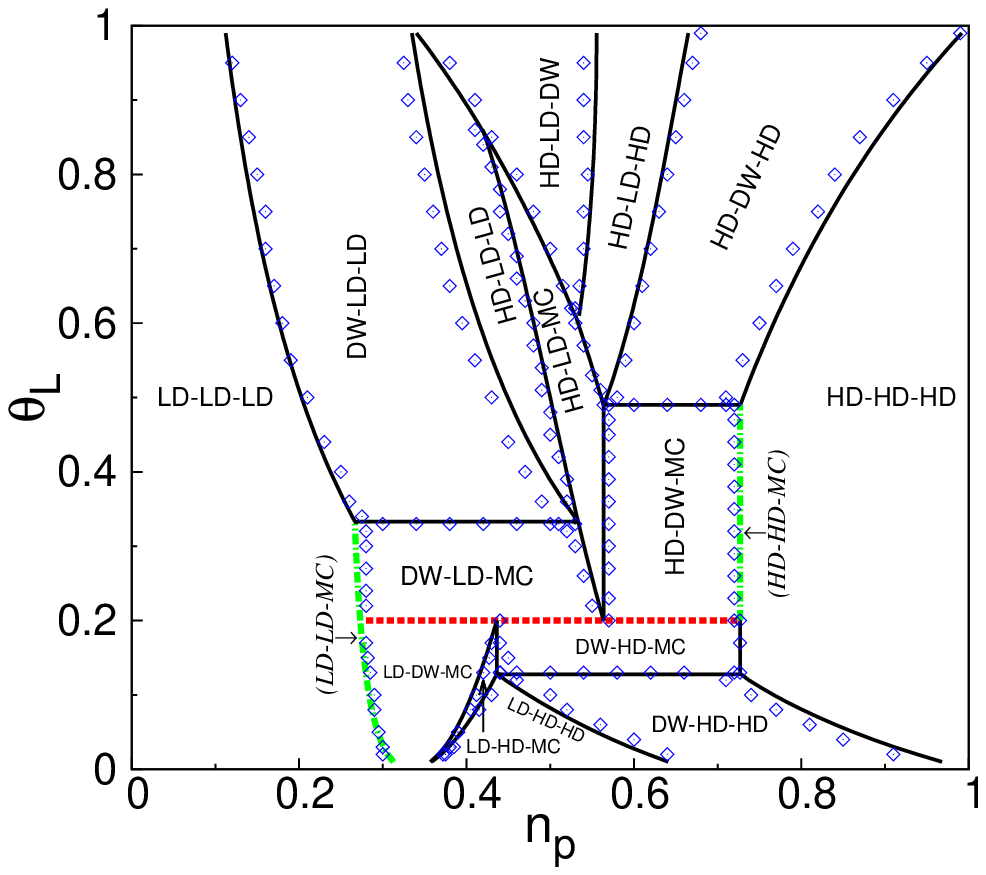}
\caption{(Color online) Phase diagram in $(n_p,\theta_L)$ -plane for
Model II with $\tr=0.20,$ phase boundaries are represented by black
continuous lines, the line-phases are denoted by green dashed-dotted
lines and DDW appears on red dotted line as obtained from MFT,
whereas corresponding data points are from MCS with $N = 500.$}
\label{phase_mft_2}
\end{figure}

The phase diagram for Model II is shown for $\tr=0.20.$ This is quite
different from the
phase diagram of Model I  in
Fig.~\ref{phase_mft} regarding the structure and the total number of
phases, as new phases emerge in comparison with Model I.

\section{Density profiles for Model II}
\label{model2density}

\subsection{$T_A$ in LD-HD coexisting phase, $T_B$ and $T_C$ both in LD phase}
\label{sec:dwldld}
\begin{figure}[h]\centering
\includegraphics[width=7cm]{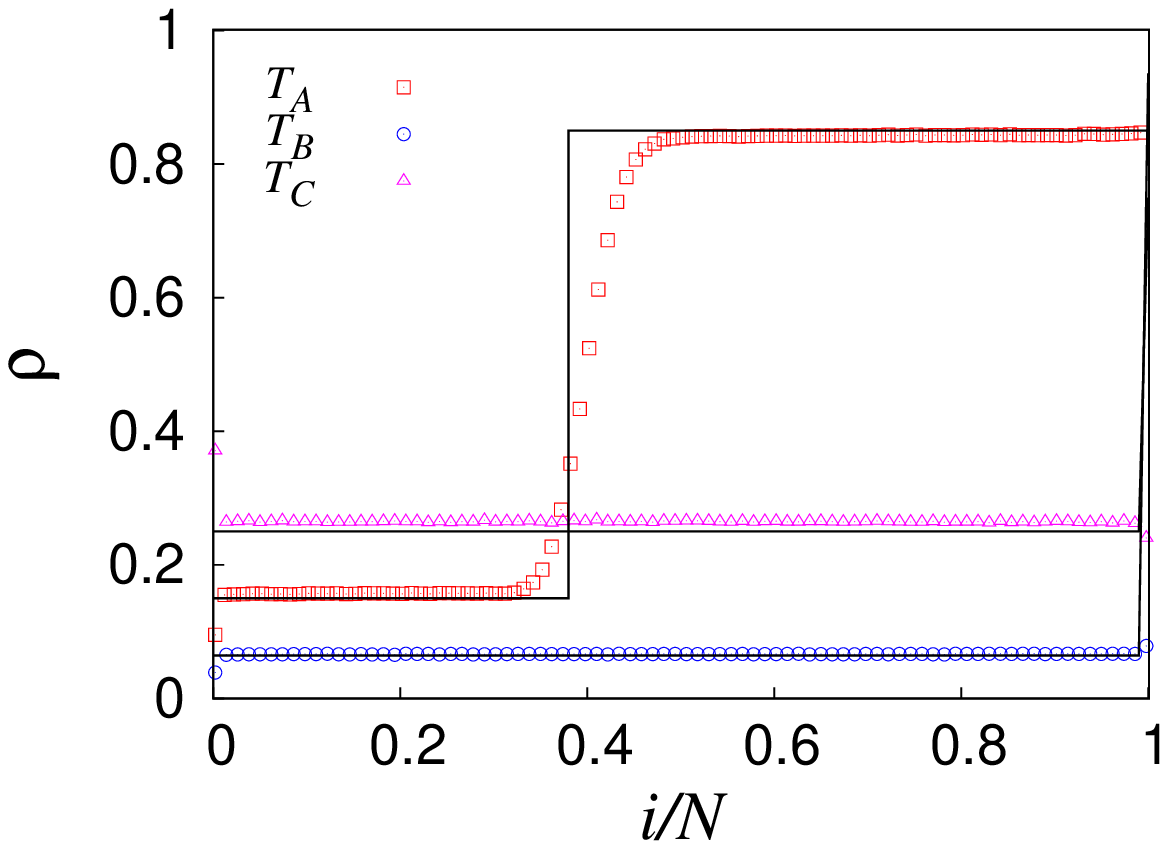}
\caption{(Color online) Density profile for the phase DW-LD-LD with $n_p=0.30,\theta_L=0.70,\tr=0.20.$}
\label{fig:DW-LD-LD}
\end{figure}

Here, $\rho_B=\alpha_B,\,\rho_C=\alpha_C$ in the bulk,
$\rho_A=\aa$
near $LJ$ and $\rho_A=1-\ba$ near $RJ$ with $\aa=\ba$, which meet
in the bulk to form an LDW; see Fig.~\ref{fig:DW-LD-LD}.  At $LJ$ and $RJ$ the
current
conservation yields,
\begin{eqnarray}
 \frac{\alpha_A}{\alpha_B}=\frac{\theta_L}{1-\theta_L},\nonumber \\
 \ba(1-\ba)=(1-\ba)(1-\ac)\tr. 
  \label{dwldld-2}
\end{eqnarray}

Now particle number conservation yields (neglecting BLs),
\begin{equation}
 3n_p=\aa+(1-x_A^w)(1-\aa-\ba)+\ab+\ac,
 \label{dwldld-5}
\end{equation}
where $x_A^w$ is the position of the LDW ($0 < x_A^w < 1$) in $T_A$. From
Eqs.~(\ref{dwldld-2}-\ref{dwldld-5} $\alpha_A=\beta_A,\alpha_B,\alpha_B$ may be
solved.
The boundaries between the LD, LD-HD phases and LD-HD, HD phases of
$T_A$ (with $T_B$ and $T_C$  in their LD phases) are obtained by
setting $x_A^w=0$ and $x_A^w=1,$ respectively in
Eq.(\ref{dwldld-5}), as shown in Fig.~\ref{phase_mft_2} for $\tr=0.20$.

\subsection{$T_A$ in LD-HD coexisting phase, $T_B$ in LD and $T_C$ in MC phase}
In this case, in the bulk $\rho_C=1/2$, $\rho_B=\ab$;
$T_A$ continues to have an LDW at $x_A^w$ as before.  Use now current
conservation to write,
\begin{eqnarray}
\frac{\alpha_A}{1-\alpha_A}=\frac{\theta_L}{1-\theta_L},\nonumber\\
 \aa(1-\aa)+\ab(1-\ab)=1/4.
 \label{dwldmc-1}
\end{eqnarray}
\begin{figure}[h]
\centering
\includegraphics[width=7cm]{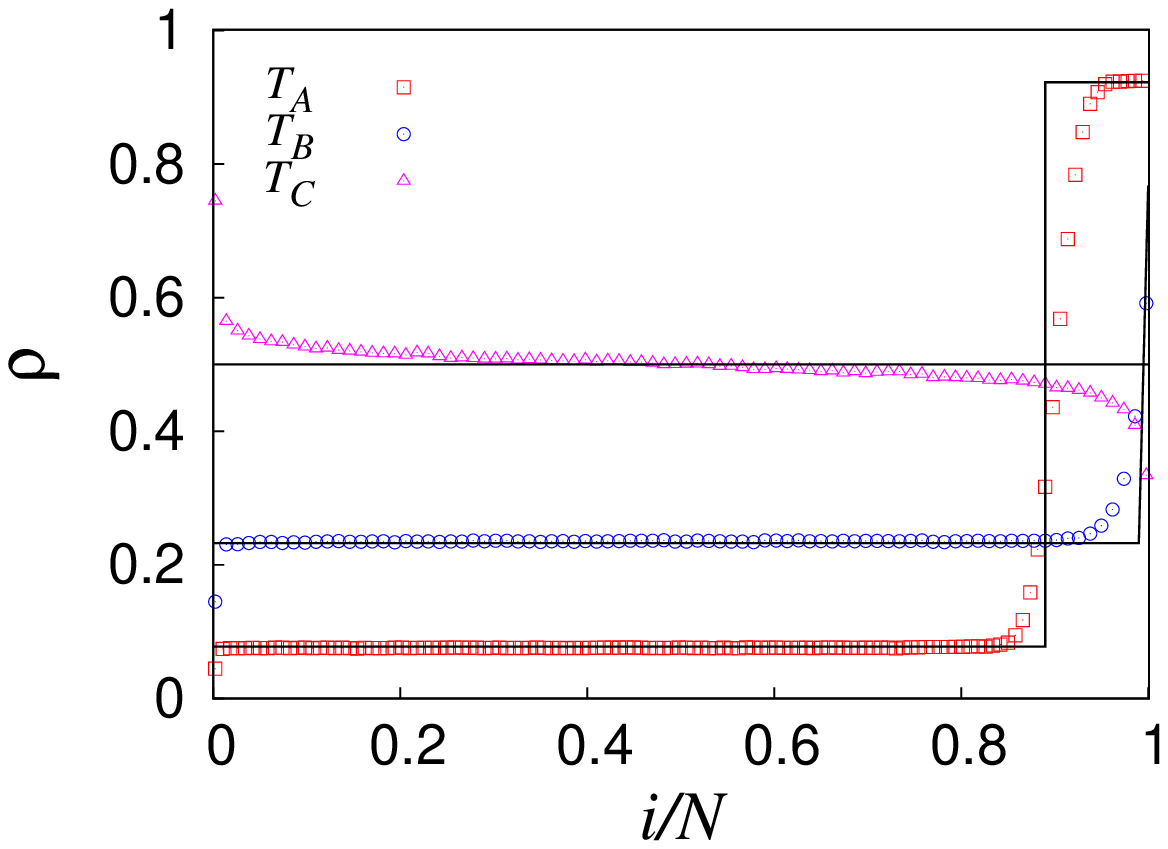}
\caption{(Color online) Density profile for the phase DW-LD-MC with $n_p=0.30,\theta_L=0.25.$}
\label{fig:DW-LD-MC}
\end{figure}

Now from particle number conservation we have,
\begin{equation}
 3n_p=\aa+(1-x_A^w)(1-\aa-\ba)+\ab+1/2.
 \label{dwldmc-4}
\end{equation}

Equations (\ref{dwldmc-1},\ref{dwldmc-4}) yield $\alpha_A,\alpha_B$ and $x_A^w$.
The phase boundaries between LD, LD-HD and LD-HD, HD phases of $T_A$ are obtained by
setting $x_A^w=0$ and $x_A^w=1$ respectively with $T_B$ in LD and $T_C$ in MC
phase,
and are shown in the phase diagram Fig.~\ref{phase_mft_2}.

\subsection{$T_A$ in LD-HD coexisting phase, $T_B$ and $T_C$ both in HD phase}
In this case $\rho_B=1-\bb$ and $\rho_C=1-\bc$ in the bulk,  and $T_A$ has an LDW.
Thus $\aa=\ba$. From current conservation
\begin{eqnarray}
 \frac{\beta_A}{\beta_B}=\frac{\theta_R}{1-\theta_R},\nonumber \\
 \aa(1-\aa)=(1-\aa)\theta_L(1-\bc).
 \label{dwhdhd-2}
\end{eqnarray}
\begin{figure}[h]\centering
\includegraphics[width=7cm]{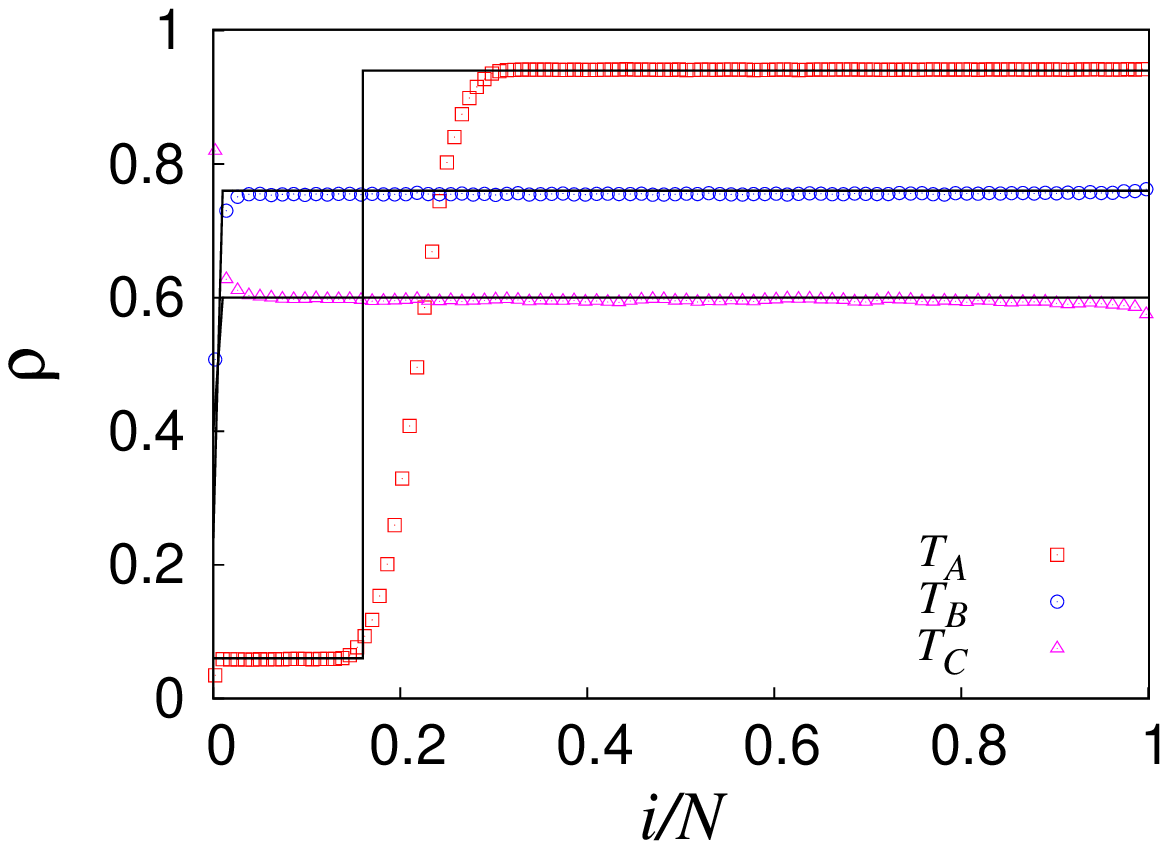}
\caption{(Color online) Density profile for the phase DW-HD-HD with $n_p=0.70,\theta_L=0.10,\tr=0.20.$}
\label{fig:DW-HD-HD}
\end{figure}

Particle number conservation yields
\begin{equation}
 3n_p=\aa+(1-x_A^w)(1-\aa-\ba)+1-\bb+1-\bc.
 \label{dwhdhd-5}
\end{equation}
where $0<x_A^w<1$ is the position of the LDW in $T_A$. Eqs. (\ref{dwhdhd-2}-\ref{dwhdhd-5}) yield $\beta_C, \aa$ and $x_A^w$. The phase boundaries
between LD,LD-HD and LD-HD,HD phases of $T_A$ are, as usual,
obtained by setting $x_A^w=0$ and $x_A^w=1$ respectively with both
$T_B$ and $T_C$ are in their HD phases. For $\tr=0.20$ the boundaries are shown in Fig.~\ref{phase_mft_2}.

\subsection{$T_A$ in LD-HD coexisting phase, $T_B$ in HD and $T_C$ in MC phase}
Here
$\rho_C=1/2$ in the bulk. Current conservation at $RJ$ gives,
\begin{eqnarray}
\frac{\beta_A}{\beta_B}=\frac{\theta_R}{1-\theta_R}.
 \label{dwhdmc-1}
\end{eqnarray}
\begin{figure}[h]\centering
\includegraphics[width=7cm]{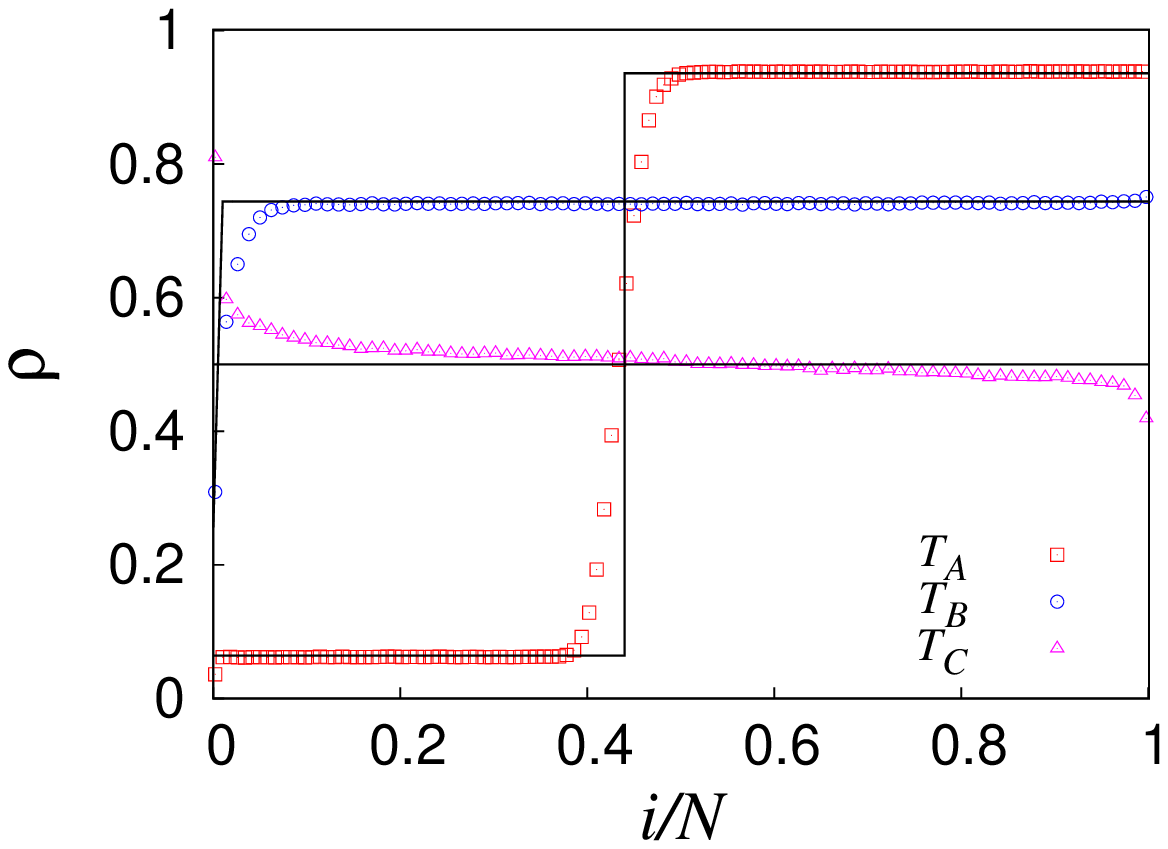}
\caption{(Color online) Density profile for the phase DW-HD-MC with $n_p=0.60,\theta_L=0.15,\tr=0.20.$}
\label{fig:DW-HD-MC}
\end{figure}

From particle number conservation we have,
\begin{equation}
 3n_p=\aa+(1-x_A^w)(1-\aa-\ba)+1-\bb+1/2.
 \label{dwhdmc-4}
\end{equation}
From Eq. \ref{dwhdmc-1} and \ref{dwhdmc-4} we $\aa,\bb$ and $x_A^w$.
The phase boundaries between LD,LD-HD and LD-HD,HD phases  are obtained by
setting
$x_A^w=0$ and $x_A^w=1$ respectively with $T_B$ in HD and $T_C$ in MC phase, and for $\tr=0.20$ the boundaries are shown in Fig.~\ref{phase_mft_2}.

\subsection{$T_A$ in HD, $T_B$ in LD-HD coexisting phase and $T_C$ in HD phase}
In this case  $\rho_A=1-\ba$ and $\rho_C=1-\bc$, respectively. Since $T_B$ is
assumed to be in its  LD-HD coexistence phase thus $\ab=\bb$.
From the current conservation we have,
\begin{eqnarray}
 \ab=(1-\bc)(1-\theta_L), \nonumber \\
 \frac{\beta_A}{\beta_B}=\frac{\theta_R}{1-\theta_R}.
 \label{hddwhd-1}
\end{eqnarray}
\begin{figure}[h]\centering
\includegraphics[width=7cm]{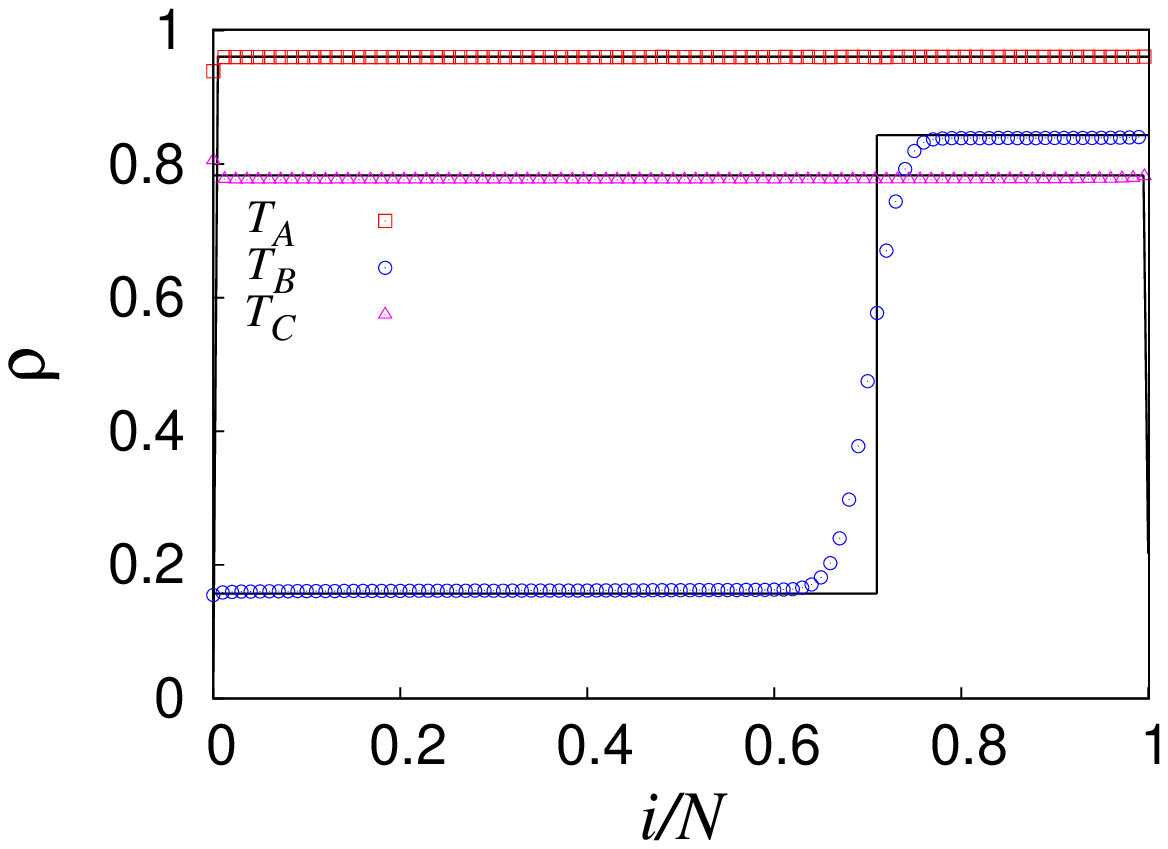}
\caption{(Color online) Density profile for the phase HD-DW-HD with $n_p=0.70,\theta_L=0.80,\tr=0.20.$}
\label{fig:HD-DW-HD}
\end{figure}

Again from particle number conservation,
\begin{equation}
 3n_p=1-\ba+\ab+(1-x_B^w)(1-\ab-\bb)+1-\bc.
 \label{hddwhd-4}
\end{equation}
Here $0<x_B^w<1$ is the position of the localised DW in $T_B$,
while $T_A$ and $T_C$ both are in HD phase. Eq.~(\ref{hddwhd-1}-\ref{hddwhd-4}) yield $\ab, \bc$ and $x_B^w$. The phase boundaries are
obtained by setting $x_B^w=0$ and $x_B^w=1,$ and for $\tr=0.20$ the boundaries are
shown in Fig.~\ref{phase_mft_2}.

\subsection{$T_A$ in HD, $T_B$ in LD and $T_C$ in LD-HD coexisting phase}
In this case
$\rho_A=1-\ba$ and $\rho_B=\ab$ in the bulk,
and $T_C$ is in DW phase. Thus $\ac=\bc$,  and $1-\bc$ give
the bulk densities on the entry ($RJ$) and exit ($LJ$) sides of $T_C$.
Now the current matching conditions at $LJ$ and $RJ$ give,
\begin{eqnarray}
\alpha_B=(1-\beta_C)(1-\theta_L),\nonumber \\
\beta_A=\theta_R(1-\alpha_C).
 \label{hdlddw-1}
\end{eqnarray}
\begin{figure}[h]\centering
\includegraphics[width=7cm]{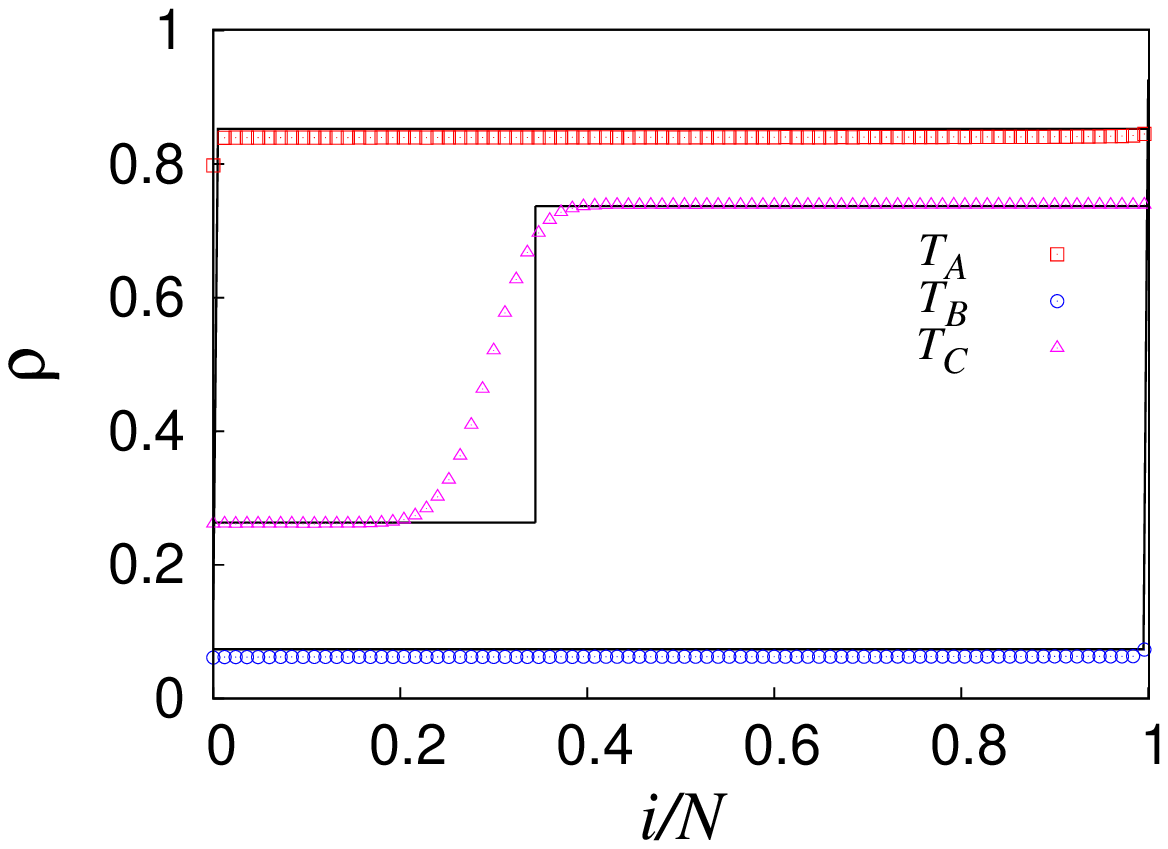}
\caption{(Color online) Density profile for the phase HD-LD-DW with $n_p=0.50,\theta_L=0.90,\tr=0.20.$}
\label{fig:HD-LD-DW}
\end{figure}

Now from particle number conservation we have,
\begin{equation}
 3n_p=1-\ba+\ab+\ac+(1-x_C^w)(1-\ac-\bc).
 \label{hdlddw-4}
\end{equation}
where $0<x_C^w<1$ is the position of the DW in $T_C$. From Eq. \ref{hdlddw-1} and \ref{hdlddw-4} we get $\ac$, $\ab$ and $x_C^w$. The phase
boundaries between LD,LD-HD and LD-HD,HD phases of $T_C$ are
obtained by setting $x_C^w=0$ and $x_C^w=1$ respectively, and for $\tr=0.20$ the
boundaries are shown in Fig.~\ref{phase_mft_2}.

\subsection{Delocalised domain walls}
\begin{figure}[h]\centering
\includegraphics[width=7cm]{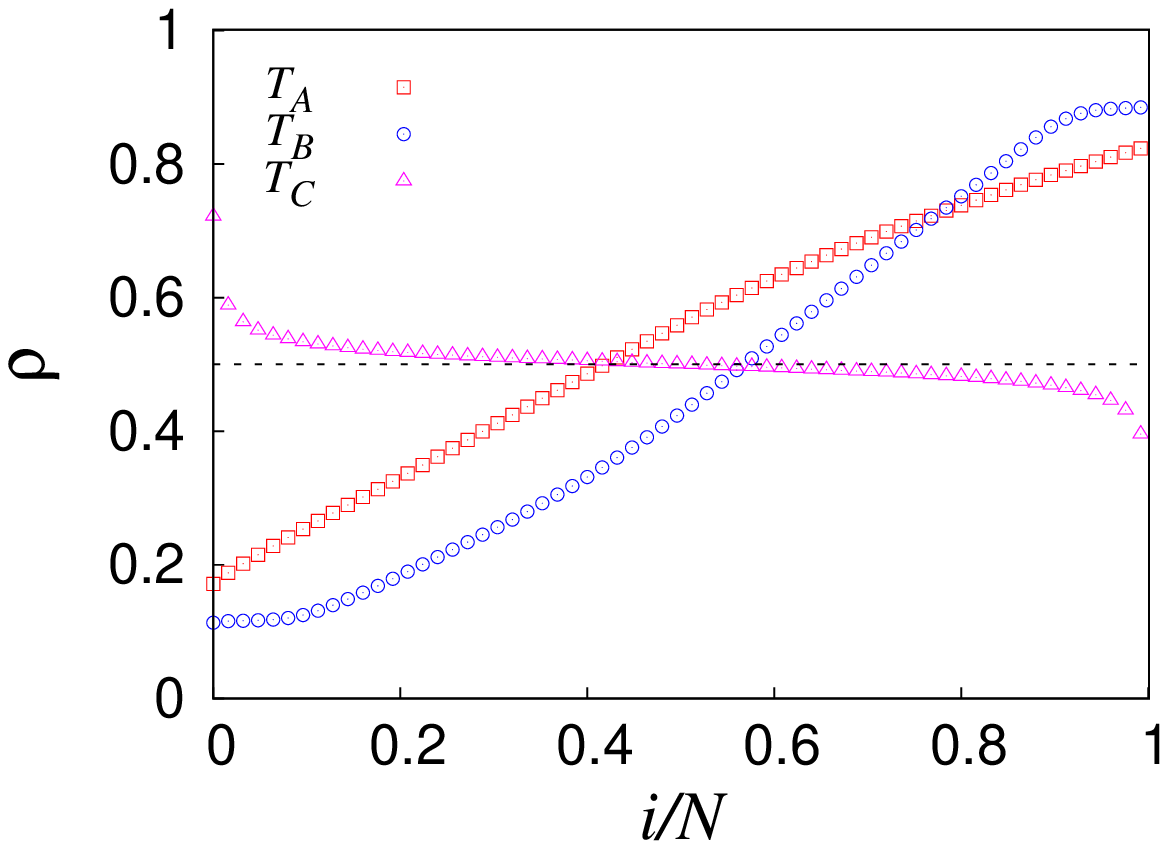}
\caption{(Color online) Density profiles when both $T_A$ and $T_B$ has DDWs with $n_p=0.50,\theta_L=0.20 = \tr$ and $N=500.$}
\label{ddw_m2}
\end{figure}

Now, assume both $T_A$ and $T_B$ have domain walls.
Then,we must have $\aa=\ba$ and $\ab=\bb.$ Again from
current matching conditions at $LJ$ and $RJ$ we have,
$\aa/\ab=\theta_L/(1-\theta_L)$ and $\ba/\bb=\tr/(1-\tr)$
respectively. These gives the condition for having domain walls in
both channels as, \be \theta_L=\tr. \label{ddw2} \ee By the similar
argument as given for Model I we can show that, both the domain
walls are delocalised with the sum of their positions  remain stable
and in that case $T_C$ remains in the MC phase. The density profiles
for the channels are shown in Fig.~\ref{ddw_m2}, while the DDW
boundary line is shown by red dotted line in phase diagram
Fig.~\ref{phase_mft_2}. Note that the DDWs in $T_A$ and $T_B$ are
non-overlapping.

\subsection{Boundary between DW-LD-LD and DW-LD-MC phases}
Here $\ac=1/2$ at this phase boundary. Now
from Eq.(\ref{dwldld-2}) we have $\aa=\tr/2$ and using the overall current conservation 
the solutions for $\theta_L$ are given by,
\begin{equation}
\theta_L=\frac{\tr^2}{\tr^2+\tr \pm
\sqrt{\tr^3(2-\tr)}}.
\label{dwldld-dwldmc-1}
\end{equation}
Clearly, $\theta_L$ is independent of $n_p$;
for $\tr=0.20$ the boundary line is
shown in the phase diagram Fig.~\ref{phase_mft_2}.

\subsection{Boundary between DW-HD-MC and DW-HD-HD phases}
At the boundary $1-\bc=1/2.$ Again from Eq.(\ref{dwhdmc-1}) we have
$\aa=\theta_L/2$ and $\bb = q\aa$, where $q=(1-\tr)/\tr.$
Now putting these values in overall current conservation equation we
have a quadratic equation in $\theta_L$ with the solutions given as,
\begin{equation}
\theta_L=\frac{1+q\pm\sqrt{2q}}{1+q^2}.
\label{dwhdmc-dwhdhd-1}
\end{equation}
Now enforcing that $T_B$ is in its HD phase, Eq.(\ref{dwhdmc-dwhdhd-1}) yields the phase boundary as a horizontal line in ($n_p,\theta_L$)-plane. For $\tr=0.20$ it is shown in  Fig.~\ref{phase_mft_2}.

\subsection{Boundary between HD-DW-MC and HD-DW-HD phases}
At the boundary $1-\bc=1/2,$ and from Eq.(\ref{hddwhd-1}) $\ab =
(1-\theta_L)/2$ and $\ba = q^{\prime}\ab$, where $q^{\prime} =
\tr/(1-\tr)$. Now putting those values in the current
conservation equation we have,
\begin{eqnarray}
 \theta_L =
 \frac{1+q^{\prime}\pm\sqrt{2q^{\prime}}}{1+{q^{\prime}}^2}.
 \label{hddwmc-hddwhd-1}
\end{eqnarray}
Now enforcing HD phase for $T_A$, we have the phase boundary (a horizontal line)  in the
($n_p,\theta_L$)-plane, as shown for $\tr=0.20$ in Fig.~\ref{phase_mft_2}.

\subsection{Boundary between HD-LD-MC and HD-LD-HD phases}
At the transition $\bc=1/2,$
and at $LJ$ we have $\ab=(1-\theta_L)/2.$
Now from overall current conservation,
\begin{eqnarray}
4\ba(1-\ba) + (1-\theta_L)(1+\theta_L)=1.
\label{hdldmc-hdldhd-1}
\end{eqnarray}
From Eq.~(\ref{hdldmc-hdldhd-1}) and  particle number conservation
\begin{eqnarray}
6n_p=3-\theta_L+\sqrt{1-\theta_L^2}.
 \label{hdldmc-hdldhd-2}
\end{eqnarray}
Equation~(\ref{hdldmc-hdldhd-2}) gives the corresponding phase boundary in the
($n_p,\theta_L$)-plane as shown Fig.~\ref{phase_mft_2}.

\subsection{Boundary between LD-HD-MC and LD-HD-HD phases}
Set $\bc=1/2,$ for this transition.
Then at the $LJ$ $\aa=\theta_L/2.$
From overall current conservation we write,
\begin{eqnarray}
 4\bb(1-\bb) + \theta_L(2-\theta_L)=1.
 \label{ldhdmc-ldhdhd-1}
\end{eqnarray}
Particle number conservation together with Eq.~(\ref{ldhdmc-ldhdhd-1}) yield 
\begin{eqnarray}
 6n_p=2+\theta_L+\sqrt{\theta_L^2-2\theta_L}.
 \label{ldhdmc-ldhdhd-2}
\end{eqnarray}
Equation (\ref{ldhdmc-ldhdhd-2}) gives the corresponding phase boundary in
($n_p,\theta_L$)-plane as
shown in  Fig.~\ref{phase_mft_2}.

\section{Comparison between Model I and Model II}
\label{apn:compare}
We
now compare and contrast the results from Model I and Model II. This
will allow us to understand the extent to which the point defects at
$RJ$ can affect the phases. First of all, the phase diagram
Fig.~\ref{phase_mft} for Model I is symmetric about the line
$\theta_L=1/2,$ whereas there is no such symmetry in the phase
diagram Fig.~\ref{phase_mft_2} for Model II in general, due to the
point defects at $RJ.$

In Model II there are some phases which has no analogue in Model I,
for example DW-LD-LD phase. Let us consider this phase where the BLs
are at the two opposite junctions, at $RJ$ for $T_B$ and at $LJ$ for
$T_C.$ At $RJ$ the current matching conditions between $T_A$ and
$T_C$ gives,
\begin{equation}
\aa(1-\aa)=(1-\aa)(1-\ac)\tra.
\label{dwldld_m1-1}
\end{equation}
Now  use the condition for retrieving Model I from Model II
given by $\tra=\trb=1,$ then from Eq.~(\ref{dwldld_m1-1})
$\aa=(1-\ac).$ Put this in the overall current
conservation equation, $\aa(1-\aa) + \ab(1-\ab) = \ac(1-\ac)$ to
get, $\ab(1-\ab) = 0$, implying $\ab=0$ or 1, which is unfeasible.
Thus, the assumption of the existence of the phase DW-LD-LD in Model
I is not correct;
therefore, we conclude that DW-LD-LD phase does not appear
for Model I. Similar arguments show that HD-LD-LD and HD-LD-DW
phases are also absent for Model I. These observations are clearly
validated by our MCS studies on Model I, as displayed in
Fig.~(\ref{phase_mft}). Evidently, the phases which appear only in
Model II are solely due to the point defect at $RJ.$ Furthermore,
the DDWs in Model I should be fully overlapping (under long time
averages; not shown in Fig~\ref{fig:ddw}), a feature consistent with
the symmetry of Model I about $\theta_L=1/2.$ In contrast, the DDWs
in Model II are generally non-overlapping (even under long time
averages), due to the lack of any special symmetry in Model II as
seen in Fig.~\ref{ddw_m2}. These differences are connected to the
fact that in Model I by construction, DDWs in $T_A$ and $T_B$
correspond to $\aa=\ba=\bb=\ab.$ In Model II however there are no
such equalities, due to the presence of the point defect at $RJ$. 
A related consequence is that Model I
displays DDWs only for $\theta_L=1/2,$ where as in Model II, it is
possible to have DDWs for arbitrary $\theta_L$ so long as the
general conditions for DDWs are met. Overall, thus, the effect of
introducing point defects at $RJ$ is not only to change the
locations of the phases in the $(n_p,\theta_L)$ phase diagram
qualitatively, but also to introduce new phases which were absent in
Model I. Thus, the topology the phase diagram gets affected.

\section{Limiting cases of Model II}
 The limiting cases of Model II reveal interesting features.
For instance, when $\theta_L$ and $\theta_R$ are either $0$ or $1$
simultaneously then by construction either $T_A$ or $T_B$ is fully
blocked. The remaining system then has equal hopping rate at every
site. Thus, the average density at every site is just $n_p$.
Furthermore, if say $\theta_L$ is $1$ or $0$ with $\tr$ having a
value between zero and unity, then the junction $RJ$ effectively
serves as a point defect in an otherwise homogeneous ring executing
TASEP. For instance, consider $\theta_L\rightarrow 1,\,0<\tr<1,$
thus eliminating $T_B$ and allowing for a point defect at $RJ$,
given by a reduced hopping rate $\tr\prime<1.$ In this limit our
model is identical to the model in Ref.~\cite{lebowitz} and our
results in this limit $\theta_L\rightarrow 1$ are in agreement with
that.

\vspace{2cm} \noindent {\em Acknowledgement:-} AB wishes to thank the Max-Planck-Gesellschaft (Germany)
and Department of Science and Technology/Indo-German Science and
Technology Centre (India) for partial financial support through the
Partner Group programme (2009). A.K.C. acknowledges the financial
support from DST (India) under the SERC Fast Track Scheme for Young
Scientists [Sanction no. SR/FTP/PS-090/2010(G)].

\end{document}